\documentstyle[12pt,epsfig]{article}
\def\pup{p^{\uparrow}}
\def\Lup{\Lambda^\uparrow} 
 
\def\qup{q^\uparrow} 
\def\bfk{\mbox{\boldmath $k$}}

\def\be{\begin{equation}}
\def\ee{\end{equation}}
\def\bea{\begin{eqnarray}}
\def\eea{\end{eqnarray}}
\def\nd{\noindent}
\begin{document}
\begin{center}
{\bf Transversity and \mbox{\boldmath $\Lambda$} polarization\footnote{Talk 
delivered at the Workshop on Future Physics @ COMPASS, Sept. 26-27, 2002,
CERN}}

\vskip 0.8cm
{\sf Mauro Anselmino}
\vskip 0.5cm
{\it Dipartimento di Fisica Teorica, Universit\`a di Torino and \\
    INFN, Sezione di Torino, Via P. Giuria 1, I-10125 Torino, Italy}\\
\end{center}

\vspace{1.5cm}

\begin{abstract}
Two related issues are discussed, which might be easily explored by 
present and future COMPASS experiments. The first one deals with the new 
world of transversity, the fundamental polarized parton distribution so far 
totally unknown. The second issue concerns $\Lambda$ production in polarized 
semi-inclusive processes, with a measurement of the $\Lambda$ polarization, 
which might give novel information on distribution and fragmentation 
properties of polarized partons. In case of transverse polarization the 
detection of $\Lambda$'s gives access to a new way of measuring transversities.
Also the interesting case of $\Lambda$ polarization in {\it unpolarized}
processes is discussed.
\end{abstract}

\section{Transversity}

The transverse polarization of quarks inside a transversely polarized nucleon,
denoted by $h_1$, $\delta q$ or $\Delta_T q$, is a fundamental twist-2 
quantity, as important as the unpolarized distributions $q$ and the 
helicity distributions $\Delta q$. It is given by
\be 
h_1(x, Q^2) = q_\uparrow^\uparrow(x, Q^2) - q_\downarrow^\uparrow(x,Q^2) \>,
\ee
that is the difference between the number density of quarks with transverse
spin parallel and antiparallel to the nucleon spin \cite{rev}. 
Fig. 1 shows the three fundamental quark distributions as seen in Deep 
Inelastic Scattering.  
\begin{figure}
\begin{center}
\includegraphics[width=10.5cm]{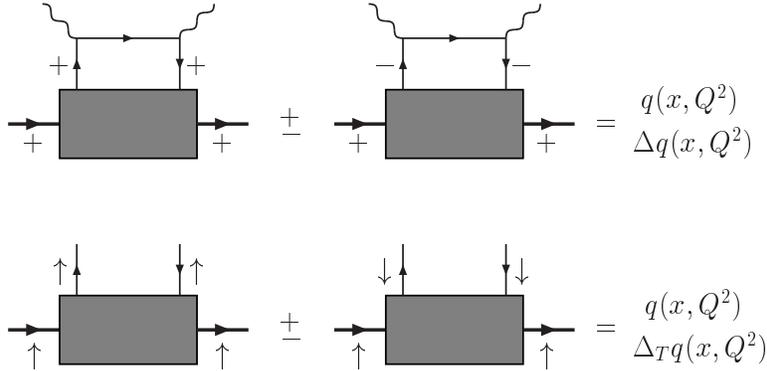} 
\caption{The three leading twist quark distributions as seen in DIS}
\end{center}
\end{figure}

Transversity is the same as the helicity distribution only in a non 
relativistic approximation, but is expected to differ from it for a 
relativistic nucleon. Not much is known bout it, apart from the fact that
it should obey the Soffer's inequality \cite{sof}
\be
2\,|h_1| \le (q + \Delta q) \,,
\ee
and that its integral is related to the tensor charge
\be
a^t_q = \int_0^1 [h_{1q}(x, Q^2) - h_{1\bar q}(x, Q^2)] \, dx \>,
\ee
for which some estimates have been obtained using non perturbative QCD
models \cite{rev}. 

When represented in the helicity basis (see Fig. 2) 
$h_1$ relates quarks with different helicities, revealing its chiral-odd 
nature. This is the reason why this important quantity has never been 
measured in DIS: the electromagnetic or QCD interactions are helicity 
conserving, there is no perturbative way of flipping helicities and $h_1$ 
decouples from inclusive DIS dynamics, as shown in Fig. 2a.
\begin{figure}
\begin{center}
\includegraphics[width=10.5cm]{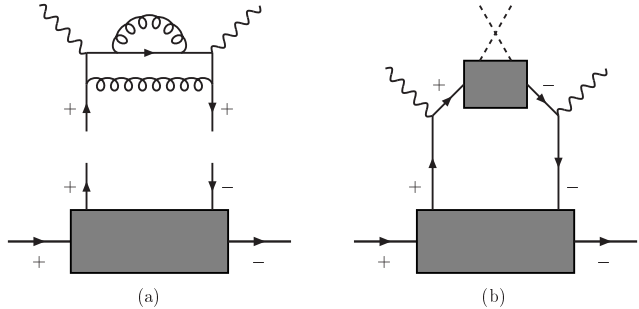} 
\caption{The chiral-odd function $h_1$ (lower box) cannot couple to inclusive
DIS dynamics, even with QCD corrections; it couples to semi-inclusive DIS,
where chiral-odd non perturbative fragmentation functions may appear.}
\end{center}
\end{figure}

However, it can be accessed in semi-inclusive deep inelastic scatterings
(SIDIS), where some non perturbative chiral-odd effects may take place in 
the non perturbative fragmentation process, Fig. 2b. Similarly, it could be 
accessed in Drell-Yan polarized processes, 
$\pup \, \pup \to \mu^+ \, \mu^- \, X$, where transverse spin asymmetries
\be
A_{TT} = \frac{d\sigma^{\uparrow\uparrow} - d\sigma^{\uparrow\downarrow}} 
{d\sigma^{\uparrow\uparrow} + d\sigma^{\uparrow\downarrow}} 
\ee
are related to the convolution of two transversity distributions. 
However, one expects very small numerical values for such asymmetries
\cite{vog}.   

\section{\mbox{\boldmath $h_1$} in SIDIS}
In order to measure the unknown transversity distribution in semi-inclusive
DIS, one needs a chiral-odd partner to associate with $h_1$; these are 
usually new fragmentation functions and several suggestions 
have been made \cite{dan}, which we shall briefly consider.   

\subsection{The Collins function}
A chiral-odd function which might occurr in the fragmentation of a 
transversely polarized quark into, say, a pion was first introduced by 
Collins \cite{col} and is schematically represented in Fig. 3; it describes
an azimuthal asymmetry in the hadronization process of a transversely
polarized quark.
\begin{figure}
\begin{center}
\includegraphics[width=10.5cm]{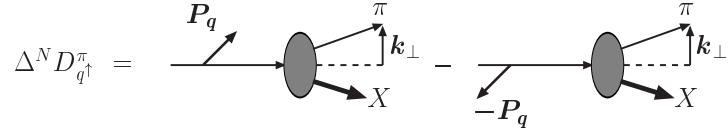} 
\caption{Pictorial representation of Collins function; notice that a similar
function is sometimes denoted by $H_1^\perp$ in the literature.}
\end{center}
\end{figure}

Such a function can give origin to a single spin asymmetry in  
$\ell \, \pup \to \ell \, \pi X$ processes, as indeed observed by HERMES 
\cite{her}, and certainly observable by COMPASS experiments:
\be
A_N = \frac{d\sigma^\uparrow - d\sigma^\downarrow}
{d\sigma^\uparrow - d\sigma^\downarrow} \> \cdot \label{an}
\ee
At leading twist, this asymmetry, if attributed to the Collins function
$\Delta^{\!N}\!D_{\pi/\qup}$, is 
given by:
\be
A^{\pi}_N =
\frac{\sum_q e_q^2 \, h_{1q}(x) \> \Delta^{\!N}\!D_{\pi/\qup}(z, k_\perp)}
{2\sum_q e_q^2 \, q(x) \> \hat D_{\pi/q}(z, k_\perp)} \>
\frac{2(1-y)} {1 + (1-y)^2} \> \sin\Phi_C \>, \label{asym1} 
\ee
where $\Phi_C$ is the azimuthal angle between the fragmenting quark 
polarization vector {\mbox{\boldmath $P_q$}} and the pion 
transverse momentum $\bfk_\perp$. Thus, $A_N$ clearly gives 
access to the transversity distributions $h_{1q}$, via the (unknown) Collins 
function: notice that a careful study of the dependence of $A_N$ on the
different DIS variables might help in obtaining separate information
on $h_1$ and $\Delta^{\!N}\!D_{\pi/\qup}$; also, selection of particular 
kinematical ranges might help in the flavour decomposition \cite{am}. 

\subsection{The Sivers function}

A mechanism similar to the Collins fragmentation was suggested for 
the proton distributions \cite{siv,noi}, and the corresponding
function denoted by $\Delta^{\!N}\!f_{q/\pup}$ or $f_{1T}^\perp$ \cite{dp};
it can again be described by Fig. 3 if one replaces the initial transversely 
polarized quark with a transversely polarized proton and the final pion 
with a quark \cite{pra}. The Sivers asymmetry was much debated, despite its 
phenomenological succes \cite{noi,bnl}, because of some supposed 
problems with QCD time-reversal properties: however,
very recently, a series of papers \cite{bro,col2,ji} have clarified the 
situation and fully promoted the rights of $\Delta^{\!N}\!f_{q/\pup}$.

When attributing the asymmetry (\ref{an}) to the Sivers mechanism, at leading 
twist, one obtains: 
\be
A^{\pi}_N =
\frac{\sum_q e_q^2 \, \Delta^{\!N}\!f_{q/\pup}(x,k_\perp) \> D_{\pi/q}(z)}
{\sum_q e_q^2 \, q(x, k_\perp) \> D_{\pi/q}(z)} \> 
\sin\Phi_S \>, \label{asym2} 
\ee
to be compared with Eq. (\ref{asym1}). The Sivers asymmetry does not allow 
access to transversity -- it is a chiral-even function -- but might 
contribute to $A_N$; such a contribution should be separated from that
of the Collins asymmetry, if we want to use data on $A_N$ to extract 
information on $h_1$. This is in principle possible if one notices that 
Eq. (\ref{asym2}) does not depend on $y$ and that the azimuthal angle 
dependence is different from the one in Eq. (\ref{asym1}); $\Phi_S$ is now
the angle between the proton polarization vector and the quark $\bfk_\perp$.   

\subsection{Other ways to approach transversity}

Other approaches to elusive transversity have been proposed \cite{dan}. 
For example, within DIS, in Ref. \cite{jaf} it was suggested to look at 
final states with two pions, originating from $s$ and $p$ wave states, 
whose interference might supply the necessary phase for a single spin 
asymmetry: these are the so-called interference fragmentation functions. 
They might avoid the danger that, in single inclusive production, the sum 
over many different channels averages the phases to zero. 

Another possibility of measuring $h_1$ goes via the SIDIS production
of spin 1 vector particles \cite{s1}; for example, one (measurable) non 
diagonal element of the helicity density matrix of a spin 1 meson, is related
to $h_1$ and some unknown fragmentation amplitudes \cite{noi2}.  

\section{\mbox{\boldmath $\Lambda$} polarization}  

Let us now turn to the second issue. 
$\Lambda$ hyperons have the peculiar feature of revealing 
their polarization through the angular distribution of their weak decay,
$\Lambda \to p \, \pi$; indeed such a feature has allowed many 
interesting measurements with unexpected and somewhat mysterious results
\cite{lpol}.

Let us consider the SIDIS processes, 
$\ell(\lambda) \, p(\mu) \to \ell \, \Lambda(h) \, X$, 
within QCD factorization theorem at leading order, with several spin 
configurations, described by the helicities $\lambda, \mu$ and $h$; 
the $\Lambda$'s are required to be produced in the current
quark fragmentation region. The kinematics and our choice of reference 
frames are explicitely shown in Fig. 4; the $\Lambda$ decay is observed 
in the helicity rest frame ($x_H, y_H, z_H$). 
\begin{figure}
\begin{center}
\includegraphics[width=10.5cm]{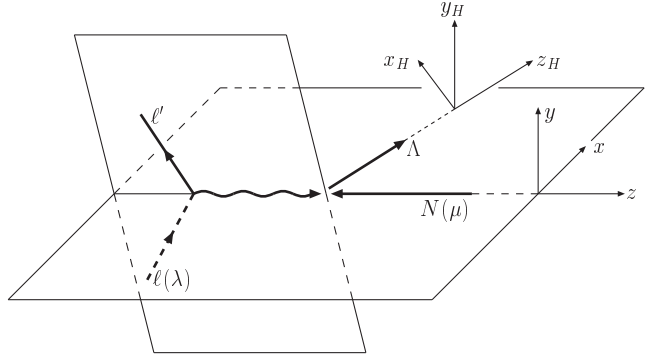} 
\caption{$\Lambda$ production in the $\gamma^* \!- p$ c.m. frame; 
the angular decay of the hyperon is measured for the particle at rest
in the helicity frame, denoted by the pedices $H$; $\lambda$, $\mu$
and $h$ denote, respectively, the initial lepton, nucleon and $\Lambda$
helicities.}
\end{center}
\end{figure}

We define
\be 
\frac{d\sigma^{\ell(\lambda) \, p(\mu)
\to \Lambda(h) \, X}}{dx \,dy \,dz} \equiv 
d\sigma_{\lambda \mu}^{\Lambda_h} \>.
\ee  
Neglecting weak interaction contributions there are 4 independent 
helicity observables, which can be chosen and written as:

\nd
the unpolarized cross-section
\be
d\sigma^{\Lambda} = 
\frac{2\pi\alpha^2}{sx} \> \frac{1 + (1-y)^2}{y^2} 
\sum_q e^2_q \, q(x) \, D_{\Lambda/q}(z) \>,
\ee
the double spin asymmetry
\be
A_{\Vert} = \frac{d\sigma_{++}^{\Lambda} - d\sigma_{+-}^{\Lambda}}
{2\,d\sigma^{\Lambda}}
=\frac{y(2-y)}{1 + (1-y)^2} \> 
\frac{\sum_q e^2_q \, \Delta q(x) \, D_{\Lambda/q}(z)}
{\sum_q e^2_q \, q(x) \, D_{\Lambda/q}(z)} \>,
\ee
the spin transfer from $\ell$ to $\Lambda$ (with an unpolarized nucleon)
\be
P_{+0} = \frac{d\sigma_{+0}^{\Lambda_+} - d\sigma_{+0}^{\Lambda_-}}
{d\sigma^{\Lambda}}
=\frac{y(2-y)}{1 + (1-y)^2} \> 
\frac{\sum_q e^2_q \, q(x) \, \Delta D_{\Lambda/q}(z)}
{\sum_q e^2_q \, q(x) \, D_{\Lambda/q}(z)} \>,
\ee
and the spin transfer from $N$ to $\Lambda$ (with an unpolarized lepton)
\be
P_{0+} = \frac{d\sigma_{0+}^{\Lambda_+} - d\sigma_{0+}^{\Lambda_-}}
{d\sigma^{\Lambda}}
= \frac{\sum_q e^2_q \, \Delta q(x) \, \Delta D_{\Lambda/q}(z)}
{\sum_q e^2_q \, q(x) \, D_{\Lambda/q}(z)} \>\cdot
\ee

The above quantities are all measurable; $P_{+0}$ means the 
polarization of the hyperon $\Lambda$ semi-inclusively produced in the DIS
scattering of a longitudinally polarized lepton (+ helicity) off
an unpolarized proton (helicity 0), and so on. These combined measurements
allow to obtain new information and/or to test available information 
on longitudinally polarized and unpolarized fragmentation and distribution 
functions, $q$, $\Delta q$, $D$ and $\Delta D$.   
A detailed discussion with numerical estimates, as well as a complete list 
of references, can be found in \cite{l1}-\cite{l3}. 
 
In particular, the above measurements should give some new information 
on the $\Lambda$ fragmentation functions; infact, from $e^+e^-$ data
one can only extract information on \cite{vog2} 
\be 
\sum_q [D_q^\Lambda + D_{\bar q}^\Lambda]  
\quad\quad {\rm and} \quad\quad
\sum_q [\Delta D_q^\Lambda - \Delta D_{\bar q}^\Lambda] \>. 
\ee 

\subsection{Transverse polarization, polarized protons}
 
We consider the process $\ell \, \pup \to \ell \, \Lup \, X$ with an 
unpolarized lepton, a transversely polarized proton ($S_N$) and the 
measurement of the $\Lambda$ trasverse polarization $P_N$; trasverse 
means orthogonal to the $\gamma^* \!-\! \Lambda$ plane, see Fig. 4. 
One has:
\be
P_N^{[0S_N]} = \frac {2(1-y)}{1 + (1-y)^2} \> \frac
{\sum_q e^2_q \, h_{1q}(x) \, \Delta_TD_{\Lambda/q}(z)}
{\sum_q e^2_q \, q(x) \, D_{\Lambda/q}(z)} \>, \label{pt}
\ee
where the transversity distribution $h_{1}$ appears coupled to 
$\Delta_T D = D_\uparrow^\uparrow -D_\uparrow^\downarrow$, 
the chiral-odd transversity fragmentation function (so far unknown). 

Eq. (\ref{pt}) offers a direct access to the product of $h_1$ and 
$\Delta_T D$ and one might hope to obtain separate information by studying 
the $x$ and $z$ dependences of $P_N$. Notice that there is no dependence
on any $k_\perp$ in this case. Notice also that neglecting 
contributions from sea quarks (which should be safe in the large $x$ and 
$z$ regions) Eq. (\ref{pt}) simplifies to:
\be
P_N^{[0S_N]} \simeq \frac {2(1-y)}{1 + (1-y)^2} \> \frac
{4 h_{1u} + h_{1d}}{4u + d} \> 
\frac{\Delta_TD_{\Lambda/u}}{D_{\Lambda/u}} \>\cdot \label{pts}
\ee

Convolutions of the same unknown functions appear in the transverse 
polarization of $\Lambda$'s produced in $p\,p$ interactions with one 
transversely polarized proton, $p \, p^\uparrow \to \Lambda^\uparrow \, X$, 
for example at RHIC \cite{vog2}:
\be
P_N(\Lambda) \sim \sum_{abc} f_{a/p} \otimes h_{1b} \otimes 
\Delta d\hat \sigma^{ab \to c\cdots} \otimes \Delta_TD_{\Lambda/c} \>,
\ee 
where $f_{a/p}$ is a parton (quark or gluon) distribution function
and the $\Delta d\hat \sigma$ are differences of polarized elementary 
QCD interactions. A combined measurement of $P_N$ in both processes might 
help to extract more information.

\subsection{Transverse polarization, unpolarized protons}

This case is particularly interesting, as it relates to the longstanding 
problem of understanding the transverse polarization of $\Lambda$'s and 
other hyperons produced in the unpolarized collisions of nucleons.
This polarization might originate from spin effects in the fragmentation 
of unpolarized quarks into polarized baryons, the so-called polarizing 
fragmentation functions \cite{abdm,mt}. These functions 
$\Delta^{\!N}\!D_{\Lup/q}$ can, again, be described by Fig. 3 if one takes an 
initial {\it unpolarized} quark and replaces the final pion with a 
transversely (up or down) polarized $\Lambda$ baryon \cite{pra}.   

Indeed the polarizing fragmentation functions can contribute to the 
transverse $\Lambda$ polarization in SIDIS \cite{noi3}; 
\bea
P_N(\Lambda,x,y,z,p_{_T}) &=&
\frac{\sum_q\,e_q^2 \, q(x)\,\Delta^{\!N}\!D_{\Lup\!/q}(z,p_{_T})}
{\sum_q\,e_q^2\,q(x)\,\hat{D}_{\Lambda/q}(z,p_{_T})}\nonumber\\
&\simeq& \frac{(4u+d)\,\Delta^{\!N}\!D_{\Lup\!/u}
+ s\,\Delta^{\!N}\!D_{\Lup\!/s}}
{(4u+d)\,\hat{D}_{\Lambda/u}+s\,\hat{D}_{\Lambda/s}} \>,
\label{ncl}
\eea
where $p_{_T}$ is the $\Lambda$ transverse momentum in the $\gamma^*\!-\!p$
c.m. frame. 

Eq. (\ref{ncl}) holds for neutral current, parity conserving, SIDIS 
processes. Even more interesting is the same quantity for the charged 
current weak process $\nu \, p \to \ell \, \Lup \, X$, investigated 
by NOMAD collaboration \cite{nom}; in such a case one has an almost direct 
measurement of the polarizing fragmentation function:
\be
P_N^{[\nu\ell]} \simeq 
\frac{\Delta^{\!N}\!D_{\Lup\!/u}}{\hat{D}_{\Lambda/u}} \>\cdot
\ee
Details and estimates can be found in Ref. \cite{noi3}.

\section{Conclusions}

The transversity distribution, last fundamental missing piece of the
polarized nucleon structure, can be accessed in semi-inclusive DIS. 
At the moment this looks like the most promising approach to the 
elusive transversity and should be strongly pursued. Ongoing and future  
COMPASS experiments offer an almost unique opportunity.

The main difficulty with measuring transversity is the necessity of 
coupling it to another unknown chiral-odd function, which is often very
interesting by itself; the actual data are always products or convolutions 
of these new functions. However, luckily, the unknown functions depend 
essentially on different kinematical variables and one can devise a strategy
to obtain separate information, provided enough data are available.       

Typically, transversity contributes to spin asymmetries; another problem 
is that of controlling other possible contributions -- independent of 
transversity -- to these asymmetries. This might arise in case of Sivers and 
Collins contributions to $A_N$; whereas the latter is coupled to $h_1$, the
former is not and such a contribution must be understood before drawing 
conclusions on $h_1$ from data on $A_N$. Also in this case some strategies
are possible.

The measurement of transverse $\Lambda$ polarization in SIDIS processes 
initiated by transversely polarized proton is a little explored, so far, 
channel to access $h_1$; the chiral-odd partner in this case is the 
transversity fragmentation function, which, again, is unknown: however,
contrary to Collins function, it does not require any intrinsic quark motion 
and does not vanish when $k_\perp = 0$. Moreover, one can expect it to be 
similar to the analogous longitudinal fragmentation function, and easy to 
model. Also, information on $\Delta_TD$ can be obtained from other processes. 

We will learn more about transversity only by combining information from
as many processes as possible, in different reactions and different 
kinematical ranges; the QCD $Q^2$ evolution of $h_1$ is known and should 
also be tested. Once a first knowledge about transversity is available, 
its many phenomenological applications to the explanation of observed spin 
asymmetries will test and improve our knowledge.

\end{document}